\documentclass[prl,reprint,twocolumn,amssymb,nofootinbib]{revtex4-1}
\usepackage{amssymb,amsmath}
\usepackage{comment}
\usepackage[retainorgcmds]{IEEEtrantools}
\usepackage{graphicx}
\usepackage{epstopdf}
\usepackage{float}
\usepackage{listings}
\usepackage{color}
\usepackage{xcolor}

\usepackage[pdftex,                
bookmarks=true,
bookmarksnumbered=true,
bookmarksopen=true, %
bookmarksopenlevel=\maxdimen,%
hypertexnames=false,
breaklinks=true,
pdfpagemode=UseOutlines,
linkbordercolor={1 0 0}]{hyperref} 

\usepackage{chngcntr}

\newcommand{\beq}{\begin{equation}}
\newcommand{\eeq}{\end{equation}\\}
\newcommand{\beqx}{\begin{equation*}}
\newcommand{\eeqx}{\end{equation*}\\}
\newcommand{\pmat}[1]{\begin{pmatrix}#1 \end{pmatrix}}

\newcommand{\rpar}[1]{\left(#1\right)}
\newcommand{\spar}[1]{\left[#1\right]}

\newcommand{\bd}{{\rm d}}

\newcommand{\TRS}{\mathcal{T}_\mathcal{RS}}
\newcommand{\TSS}{\mathcal{T}_\mathcal{SS}}
\newcommand{\dTRS}{{\dot{\mathcal{T}}_\mathcal{RS}}}
\newcommand{\ddTRS}{{\ddot{\mathcal{T}}_\mathcal{RS}}}
\newcommand{\dddTRS}{{\dddot{\mathcal{T}}_\mathcal{RS}}}

\begin{document}

\title{Running of the Running and Entropy Perturbations During Inflation}

\author{Carsten van de Bruck}
\email{C.vandeBruck@sheffield.ac.uk}
\affiliation{Consortium for Fundamental Physics, School of Mathematics and Statistics, \\
		University of Sheffield, Hounsfield Road, Sheffield, S3 7RH, United Kingdom}

\author{Chris Longden}
\email{cjlongden1@sheffield.ac.uk}
\affiliation{Consortium for Fundamental Physics, School of Mathematics and Statistics, \\
		University of Sheffield, Hounsfield Road, Sheffield, S3 7RH, United Kingdom}

\date{\today}

\begin{abstract}
In single field slow-roll inflation, one expects that the spectral index $n_s -1$ is first order in slow-roll parameters. Similarly, its running $\alpha_s = \bd n_s/\bd \log k$ and the running of the running $\beta_s = \bd\alpha_s/\bd \log k$ are second and third order and therefore expected to be progressively smaller, and usually negative. Hence, such models of inflation are in considerable tension with a recent analysis hinting that $\beta_s$ may actually be positive, and larger than $\alpha_s$. Motivated by this, in this work we ask the question of what kinds of inflationary models may be useful in achieving such a hierarchy of runnings, particularly focusing on two--field models of inflation in which the late-time transfer of power from isocurvature to curvature modes allows for a much more diverse range of phenomenology. We calculate the runnings due to this effect and briefly apply our results to assess the feasibility of finding $|\beta_s| \gtrsim |\alpha_s|$ in some specific models.
\end{abstract}

\maketitle


Constraining models of inflation is one of the most important goals of cosmology. By constraining or even ruling out models of inflation, cosmologists learn a great deal about model building in theories beyond the standard model. Even with the latest cosmological observations \cite{Ade:2015lrj,Ade:2015tva}, there is still a plethora of inflationary models compatible with data \cite{Martin:2013tda}. It was recently pointed out that observations of the cosmic microwave background (CMB) radiation are consistent with a rather large running of the running of the spectral index\footnote{We denote the running of the spectral index by $\alpha_s = \bd n_s / \bd \log k$, and its running $\beta_s = \bd \alpha_s / \bd \log k$}, \cite{Escudero:2015wba,Cabass:2016ldu}. The constraints on $\alpha_s$ and $\beta_s$ given in \cite{Cabass:2016ldu} are $\alpha_s = 0.011 \pm 0.010$ and $\beta_s = 0.027\pm0.013$ (fixing the pivot scale at $k = 0.05$Mpc$^{-1}$), which implies $\beta_s > 0$ at the $2\sigma$ confidence level and hints that the running of the running may be larger than the running itself. While albeit a weaker hint, $\alpha_s$ also appears to be positive, which leads to a slight tension with a wide range of inflationary models that predict a negative running \cite{Gariazzo:2016blm}. Future CMB experiments are needed to determine these parameters more precisely, but given this first hint of what is potentially a powerful piece of evidence in early universe cosmology, it is interesting and worthwhile to consider the theoretical viability for such a hierarchy of runnings to be realised in inflation. 

In standard single field slow--roll models, the running is of second order in slow--roll parameters and the running of the running is third order (see eqs. (\ref{eq:horizonns})-(\ref{eq:horizonbeta}) below). Thus, in such models of inflation, one would quite generally expect $\beta_s$ to be smaller than $\alpha_s$. These observational hints motivate the current work, in which we study predictions of $\alpha_s$ and $\beta_s$ in single and two--field inflationary scenarios with the intention of understanding what kinds of inflationary models could be consistent with such a hierarchy of runnings.

While almost all investigations of inflationary models make predictions for the spectral index, relatively few study the running \cite{Peloso:2014oza,Garcia-Bellido:2014gna,Gariazzo:2016blm,Kohri:2014jma}, and almost none discuss the running of the running \cite{Bojowald:2011iq,Escudero:2015wba}. This is largely understandable, given how until fairly recently we did not even have tight bounds on the more easily measurable quantities like $n_s$. However, now, as we seek to further narrow down the plethora of proposed models, alongside tests of inflation non-gaussianities and spectral distortions, it is possible that interesting constraints and physical insight could come from predictions and measurements of $\alpha_s$ and $\beta_s$, particularly if such a previously unexpected hierarchy is confirmed to exist at a higher statistical significance by future experiments. 



To be concrete, we assume that gravity is described by General Relativity but allow a general scalar field Lagrangian, $P$, depending on two fields $\phi^I$ and kinetic terms $X^{JK}=\frac{1}{2}g^{\mu\nu} (\partial\phi^J / \partial x^\mu)(\partial\phi^K / \partial x^\nu)$, ($I, J, K = 1, 2$). The action is therefore given by (we set $M_{\rm Pl} = 1$)  
\beq
S = \int \bd^4 x \sqrt{-g} \spar{\frac{1}{2} R + P(\phi^I,X^{JK})}.
\eeq
This action encompasses a wide range of models, including coupled and uncoupled two-field models \cite{Kaiser:2013sna,Schutz:2013fua}, those with Diract Born Infeld (DBI) kinetic terms \cite{Alishahiha:2004eh}, disformally coupled inflation \cite{vandeBruck:2015tna}, and even many theories with non-standard gravity following a transformation of the action (such as Starobinsky inflation after conformal transformation \cite{vandeBruck:2015xpa,Kaneda:2015jma}). 

To describe slow--roll inflation, we make use of the following slow--roll parameters, which are defined recursively by 
\beq
\epsilon_0 = -\frac{\dot{H}}{H^2} \, ,\quad \epsilon_{n+1} = \frac{\dot{\epsilon}_{n}}{H \epsilon_n} \, ,
\eeq
where $H$ is the expansion rate during inflation and the dot denotes the derivative with respect to cosmic time. These slow--roll parameter are assumed to be small \mbox{($ \epsilon_n \ll O(1)$)} and approximately constant. In this regime, the power spectrum $\mathcal{P}_\mathcal{R}$ of the curvature perturbation ${\cal R}$ at horizon crossing is given at leading order by (we use the symbol $\simeq$ to denote expressions which are valid in the slow--roll approximation) \cite{Garriga:1999vw,Langlois:2008qf}
\beq
\mathcal{P}_\mathcal{R}^* \simeq \frac{H^2}{8 \pi^2 \epsilon_0 c_s} \, ,
\eeq
where $c_s$ is the sound speed of the adiabatic perturbation, and we use an asterisk to signify the value of a quantity at the moment of horizon exit ($c_s k = a H$). We also need to define a series of slow-roll-like (in that they obey the same assumptions as the $\epsilon_n$) parameters related to $c_s$ with 
\beq
s_0 = \frac{\dot{c}_s}{H c_s} \, ,\quad s_{n+1} = \frac{\dot{s}_{n}}{H s_n} \, .
\eeq
It is straightforward to evaluate the spectral index $n_s$, its running $\alpha_s$ and the running of the running $\beta_s$ in the lowest order slow--roll approximation. One finds 
\begin{eqnarray}
\rpar{n_s^* - 1} &\equiv& \left. \frac{\bd \ln \mathcal{P}_\mathcal{R}}{ \bd \ln k}\right|_{c_s k=aH}  \simeq - 2 \epsilon_0 - \epsilon_1 - s_0~, \label{eq:horizonns}\\ 
\alpha_s^* &\equiv& \left.  \frac{\bd n_s}{\bd \ln k}\right|_{c_sk=aH} \simeq- 2 \epsilon_0 \epsilon_1 - \epsilon_1 \epsilon_2 - s_0 s_1~, \label{eq:horizonalpha}\\
\beta_s^* &\equiv& \left.  \frac{\bd \alpha_s}{\bd \ln k}\right|_{c_sk=aH} \simeq - 2 \epsilon_0 \epsilon_1 (\epsilon_1 + \epsilon_2) \nonumber \\
&-& \epsilon_1 \epsilon_2 (\epsilon_2 + \epsilon_3) -  s_0 s_1 (s_1 + s_2)~, \label{eq:horizonbeta}
\end{eqnarray}
in which the slow--roll parameters are evaluated at horizon crossing.


The above results are the final predictions for single--field models, as $\mathcal{R}$ is approximately constant outside the horizon. However, as it is well known, in theories with multiple fields entropy perturbations can source the evolution of the curvature perturbation $\mathcal{R}$ outside the horizon. Thus, it is not sufficient to calculate the spectral properties at horizon crossing and one must also take in to account the effect of isocurvature modes on superhorizon scales. To account for this, we use the transfer function formalism where the total power spectrum at the end of inflation is related to the horizon crossing spectrum by \cite{Wands:2002bn}
\beq
\mathcal{P}_\mathcal{R} = \mathcal{P}_\mathcal{R}^* \rpar{1 + \TRS^2} \equiv \frac{\mathcal{P}_\mathcal{R}^*}{\cos^2{\Theta}}\, ,
\eeq
where $\TRS$ is the transfer function encoding the growth of ${\cal R}$ due to entropy perturbations, and $\Theta = \tan^{-1} \TRS$ is the transfer angle. It is then easy to derive the spectral index {\it at the end of inflation} as follows 
\begin{eqnarray}
\rpar{n_s - 1} &=& \frac{\bd \ln \mathcal{P}_\mathcal{R}^*}{\bd \ln k} + \frac{\bd \ln \rpar{1+\TRS^2}}{ \bd \ln k} \nonumber \\
&\simeq & \rpar{n_s^* - 1} + \frac{1}{H^*} \frac{\bd \ln \rpar{1+\TRS^2}}{\bd t_*}  \, , \label{eq:fullns}
\end{eqnarray}
where we have decomposed the result into the part depending only on the power spectrum at horizon crossing (i.e. $n_s^*$ is given by eq.(\ref{eq:horizonns})), and the part representing corrections due to isocurvature perturbations. One finds for $\alpha_s$ and $\beta_s$ at the end of inflation  
\begin{eqnarray}
\alpha_s & \simeq &  \alpha_s^* + \frac{1}{H^2_*} \frac{\bd^2 \ln \rpar{1+\TRS^2}}{\bd t_*^2} \, , \\
\beta_s & \simeq & \beta_s^* +  \frac{1}{H^3_*} \frac{\bd^3 \ln \rpar{1+\TRS^2}}{\bd t_*^3} \, \label{eq:fullbeta}
\end{eqnarray}
with $\alpha_s^*$ and $\beta_s^*$ are given by eqns. (\ref{eq:horizonalpha}) and (\ref{eq:horizonbeta}), respectively. 

To understand this further, we need to use some results from the transfer function formalism \cite{Wands:2002bn}. Generally, on superhorizon scales, the feeding of curvature perturbations $\mathcal{R}$ by entropy perturbations $\mathcal{S}$ can be modelled by equations of the form
\beq
\dot{\mathcal{R}} \simeq A H \mathcal{S} \, , \quad \dot{\mathcal{S}} \simeq B H \mathcal{S} \, ,
\eeq
where $A$ and $B$ are model-dependent couplings between adiabatic and entropy modes. The solution of these equations can be written in matrix form as 
\beq
\pmat{\mathcal{R} \\ \mathcal{S}} = \pmat{1 & \TRS \\ 0 & \TSS}\pmat{\mathcal{R} \\ \mathcal{S}}^* \, ,
\eeq
where the transfer functions are given by
\beq
\TSS(t) = \exp\rpar{\int_{t^*}^{t} B(t) H(t) \bd t} \, ,
\eeq
and
\beq \label{eq:TRSintegral}
\TRS(t) = \int_{t^*}^{t} A(t) H(t) \TSS(t) \bd t \, .
\eeq
From this latter expression we can work out the derivatives of $\TRS$ with respect to $t^*$ that appear in (\ref{eq:fullns})-(\ref{eq:fullbeta}). We find 
\begin{eqnarray}
\dTRS &\simeq& - H_* \rpar{A_* + B_* \TRS}, \nonumber \\
\ddTRS &\simeq& H_*^2 \rpar{A_*B_* + B_*^2 \TRS}, \nonumber \\
\dddTRS &\simeq& -H_*^3 \rpar{A_*B_*^2 + B_*^3 \nonumber}.
\end{eqnarray}
Using these time derivatives and using the definition of the transfer angle $\Theta$ ($\TRS = \tan \Theta$), we find 
\begin{eqnarray}
&n_s & \simeq n_s^*  - 2 \sin \Theta \rpar{A_* \cos \Theta + B_* \sin \Theta}\, ,\label{eq:ns} \\
&\alpha_s& \simeq \alpha_s^* + 2 \cos \Theta \rpar{A_* \cos \Theta + B_* \sin \Theta} \nonumber \\
& & \times \rpar{A_* \cos 2 \Theta + B_* \sin 2 \Theta} \, ,\label{eq:alpha}\\
&\beta_s& \simeq \beta_s^*  - 2 \cos \Theta \rpar{A_* \cos \Theta + B_* \sin \Theta}  \nonumber \\ 
& & \times \rpar{B_* \cos 2 \Theta - A_* \sin 2 \Theta}\nonumber \\
& & \times \rpar{A_* + 2A_* \cos 2\Theta + 2 B_* \sin 2 \Theta}\, .\label{eq:beta}
\end{eqnarray}
We have hence obtained expressions for the spectral index and runnings in a general two-field model of inflation that depend on quantities evaluated at horizon exit ($n_s^*$, $\alpha_s^*$, $\beta_s^*$, $A_*$ and $B_*$) which are relatively easy to compute, and one variable which parametrises our ignorance of the more involved superhorizon evolution of perturbations, the transfer angle $\Theta$. Note that the factor $\rpar{A_* \cos \Theta + B_* \sin \Theta}$ appears in all three expressions (\ref{eq:ns})-(\ref{eq:beta}) and should hence not be small if we are to have a large $\beta_s$, assuming that $\beta_s^*$ is negligible, which we have argued is expected. To suppress $\alpha_s$ while allowing $\beta_s$ to remain potentially large it is the factor $\rpar{A_* \cos 2 \Theta + B_* \sin 2 \Theta}$ in $\alpha_s$ which, if made small, would most readily facilitate this. 

Finally, for completeness we give the consistency relation between the tensor-to-scalar ratio $r$, the tensor spectral index $n_T$, the sound speed of adiabatic perturbations $c_s$ and the transfer angle, given by $\Theta$ \cite{Wands:2002bn} 
\begin{equation}\label{eq:twofieldcr}
r \simeq -8 n_T c_s \cos^2 \Theta.
\end{equation}
This consistency relation is particularly useful in the context of this work in that it relates the tensor power spectrum to the isocurvature transfer angle, which we have shown in (\ref{eq:ns})-(\ref{eq:beta}) to influence the values of $\alpha_s$ and $\beta_s$. Combining information on the running of the running with information on the tensor spectrum will further strengthen our capability to constrain and test models of inflation.


Having derived the expressions for $n_s$, $\alpha_s$ and $\beta_s$, we now discuss whether it is achievable to obtain $\alpha_s \approx \beta_s$ or even $\beta_s > \alpha_s$ from an inflationary model. In single field inflation, the predictions for the spectral properties are given by eqns. (\ref{eq:horizonns}), (\ref{eq:horizonalpha}) and (\ref{eq:horizonbeta}). The models discussed in \cite{Escudero:2015wba}, in which $c_s = 1$ and hence $s_i = 0$ for $i\geq0 $, predict negative values for $\beta_s$. The only way for having $\alpha_s$ and $\beta_s$ of the same order of magnitude, while maintaining an acceptably small $n_s -1$, is by making the second and third term in eq. (\ref{eq:horizonbeta}) relatively large and having the right sign to make $\beta_s>0$. This implies either a relatively large $\epsilon_3$ or $s_2$ (or both). This can be achieved in models which violate slow-roll, such as those in which the potential has features, for example, if the first and second derivatives of $\epsilon$ and $c_s$ are small and only higher derivatives are large (the type of models studied in\cite{Adams:2001vc,Ashoorioon:2008qr,Ashoorioon:2014yua} are not in this class of models). This would likely require fine tuning of the coefficients of the lower order terms in the effective potential. Alternatively, in K--essence models \cite{ArmendarizPicon:2000ah} the kinetic term would need to have specific properties such that $s_2$ becomes (relatively) large at horizon crossing, but $s_1$ remains relatively small. All of this has to be done in such a way that $n_s-1$ as well as $\alpha_s$ remain small and only $\beta_s$ is made relatively large. While it appears possible to build such a model, it is probably not very natural in the setup we are considering. Thus we find that single field inflationary models generally predict the hierarchy $|n_s-1 |> |\alpha_s| > |\beta_s|$. We therefore turn our attention to a two--field model, in which isocurvature modes affect the final values of $n_s$, $\alpha_s$ and $\beta_s$.  

\begin{figure}[t]
    \centering
    \includegraphics[scale=0.425]{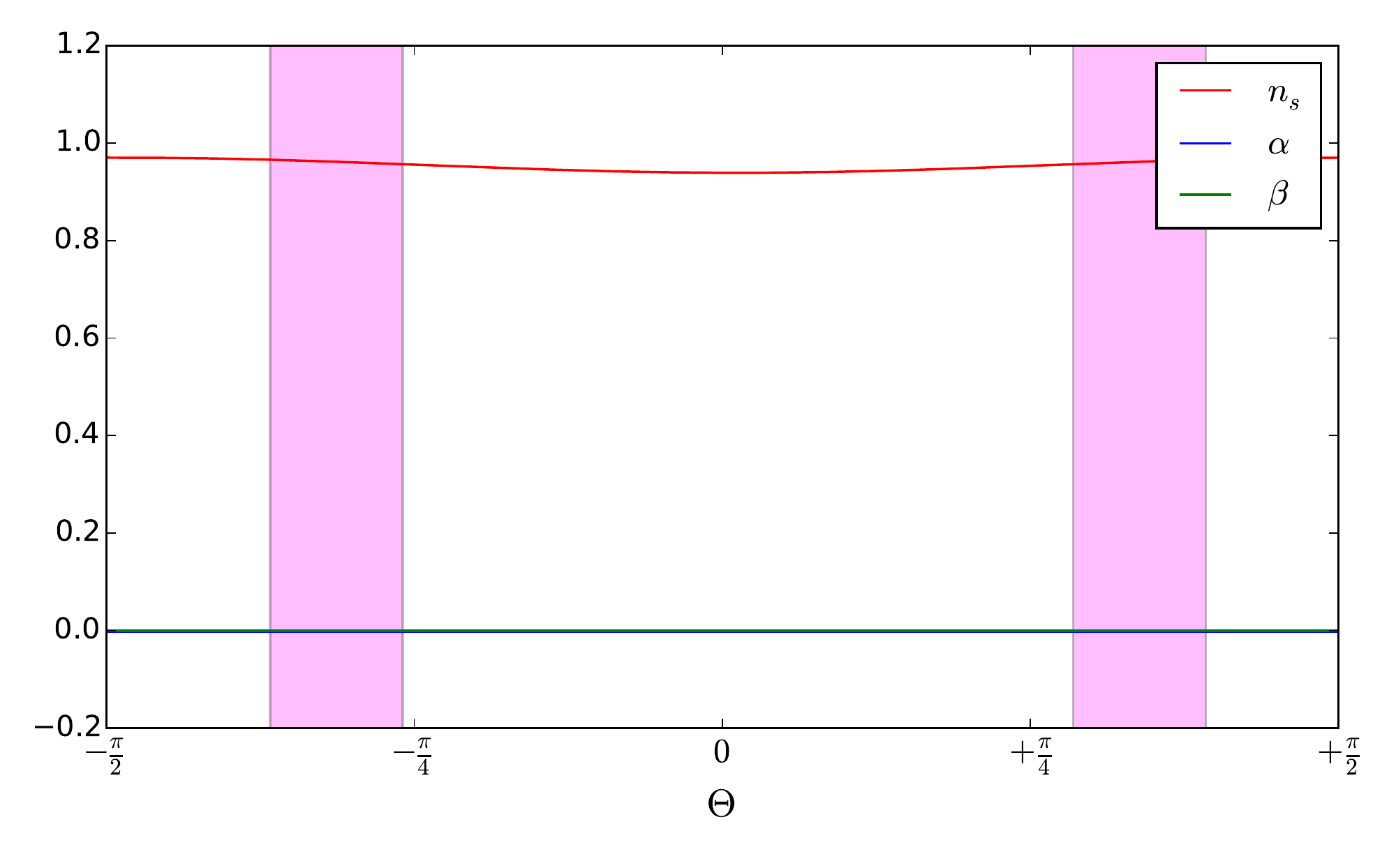}
    \includegraphics[scale=0.425]{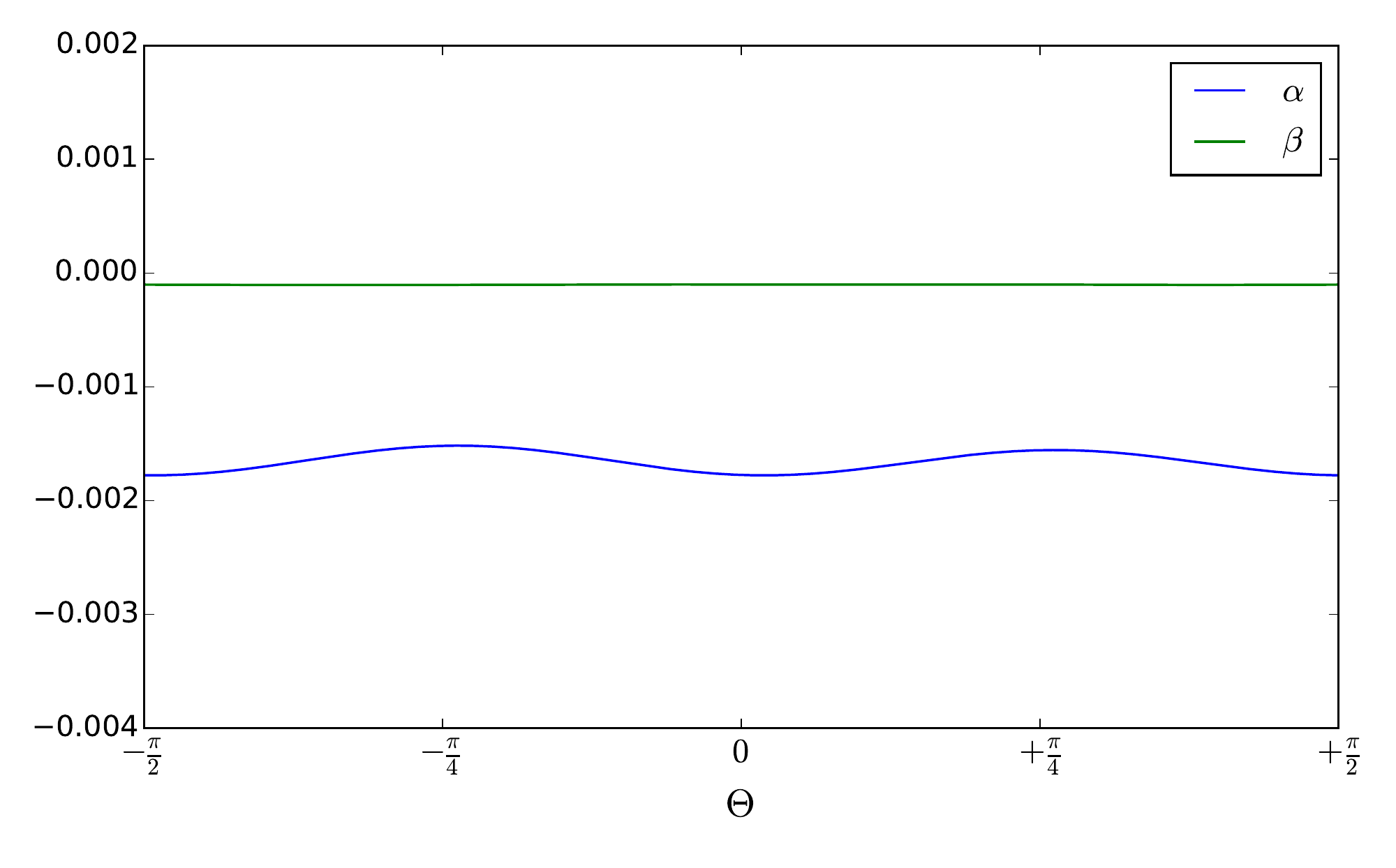}
    \caption{Predictions for the case (\ref{caseone}), with $m = 5 \times 10^{-6}M_{\rm Pl}$, $M = 1\times 10^{-6}M_{\rm Pl}$. The upper plot shows the predictions for $n_s$, $\alpha_s$ and $\beta_s$ as a function of the transfer angle $\Theta$, the lower plot shows the predictions for $\alpha_s$ and $\beta_s$ in more detail. The shaded region shows the values for $\Theta$ for which $n_s$ lies in the measured range. This model predicts negative values for $\alpha_s$ and $\beta_s$ with $|\alpha_s|>|\beta_s|$. We emphasise that this qualitative behaviour is typical of this model, and in particular is independent of the mass ratio of the two fields.}
    \label{fig:one}
\end{figure}

To be specific, we are looking at models of the type \cite{DiMarco:2002eb,DiMarco:2005nq,Lalak:2007vi,vandeBruck:2014ata}
\beq
S = \int \bd^4 x \sqrt{-g} \spar{\frac{1}{2} R + X^{\phi\phi} + e^{2 b(\phi)} X^{\chi\chi} - V(\phi,\chi)} \, ,
\eeq
in which $c_s = 1$ (and hence $s_n = 0$). For this model, one can easily find expressions for $A$ and $B$. They are given by \cite{DiMarco:2005nq} 
\begin{eqnarray}
A &\simeq& -2 \eta_{\sigma s} - \epsilon_{b\chi} \sin^2 \theta \, , \label{eq:A}\\
B &\simeq& (\eta_{\sigma\sigma} - \eta_{ss}) - 2 \epsilon_0 \nonumber \\
 &-& \frac{1}{2} \epsilon_{b\chi} \rpar{1 + \sin^2 \theta - \sin \theta \cos \theta} \label{eq:B}\, ,
\end{eqnarray}
where $\cos \theta = \dot{\phi}/\dot{\sigma}$, $\sin \theta = e^{b}\dot{\chi}/\dot{\sigma}$ with $\dot{\sigma}^2 = \dot{\phi}^2 + \dot{\chi}^2 e^{2b}$ and 
\begin{eqnarray}
\epsilon_{b\chi} &=& 2 \frac{V_{,\chi} b_{,\phi}}{V}, \nonumber \\
\eta_{\sigma\sigma} &=& \frac{V_{,\phi\phi}}{V} \cos^2 \theta + \frac{V_{,\chi\chi}}{V} e^{-2 b(\phi)} \sin^2 \theta + \frac{V_{,\phi\chi}}{V} e^{- b(\phi)} \sin 2\theta \nonumber \\
\eta_{ss} &=& \frac{V_{,\phi\phi}}{V} \sin^2 \theta + \frac{V_{,\chi\chi}}{V}e^{-2 b(\phi)} \cos^2 \theta - \frac{V_{,\phi\chi}}{V} e^{- b(\phi)} \sin 2\theta\nonumber \\
\eta_{\sigma s} &=& \frac{(V_{,\chi\chi} e^{-2 b(\phi)} - V_{,\phi\phi})}{V} \sin \theta \cos \theta + \frac{V_{,\phi\chi}}{V} e^{- b(\phi)} \cos 2 \theta.\nonumber \\
\end{eqnarray}
Inspecting eqns. (\ref{eq:ns})--(\ref{eq:beta}), we would like the values of $A$ and $B$ at horizon crossing to be somewhat large to achieve a hierarchy such as $|\beta_s| >  |\alpha_s|$.

We first look at the simplest case of two non-interacting massive scalar fields with no kinetic coupling, that is,
\beq \label{caseone}
V = \frac{1}{2} m^2 \phi^2 + \frac{1}{2} M^2 \chi^2 \, , \quad b = 0 \, .
\eeq
For this case, we find that 
\begin{align}
A & \simeq\frac{4 (1-R^2)}{(\phi^2+\chi^2)(\phi^2+R^2 \chi^2)} \dot{\phi}\dot{\chi} \, , \\
B & \simeq -2 \epsilon_0 + \frac{2 (1-R^2)}{(\phi^2+\chi^2)(\phi^2+R^2 \chi^2)} \rpar{\dot{\phi}^2 - \dot{\chi}^2} \, ,
\end{align}
where $R= M/m$. For similar masses ($R \approx 1$), both $A$ and $B$ are close to zero at horizon crossing. When one mass is much larger than the other ($R \rightarrow 0$ or $R \rightarrow \infty$), both terms are still slow-roll suppressed. Thus, $A$ and $B$ are small in this model. As a result, we find that isocurvature modes cannot break the hierarchy $|n_s -1| > |\alpha_s| > |\beta_s|$ irrespective of the details of the large-scale evolution encoded in $\Theta$, and in particular, $\beta_s$ will remain small. We show the results for $n_s$, $\alpha_s$ and $\beta_s$ in this model, using eqns. (\ref{eq:ns})--(\ref{eq:beta}), in Fig. \ref{fig:one}. We find that the situation shown in Fig \ref{fig:one} is typical for this model and it meets our analytical expectations.

\begin{figure}[t]
    \centering
    \includegraphics[scale=0.425]{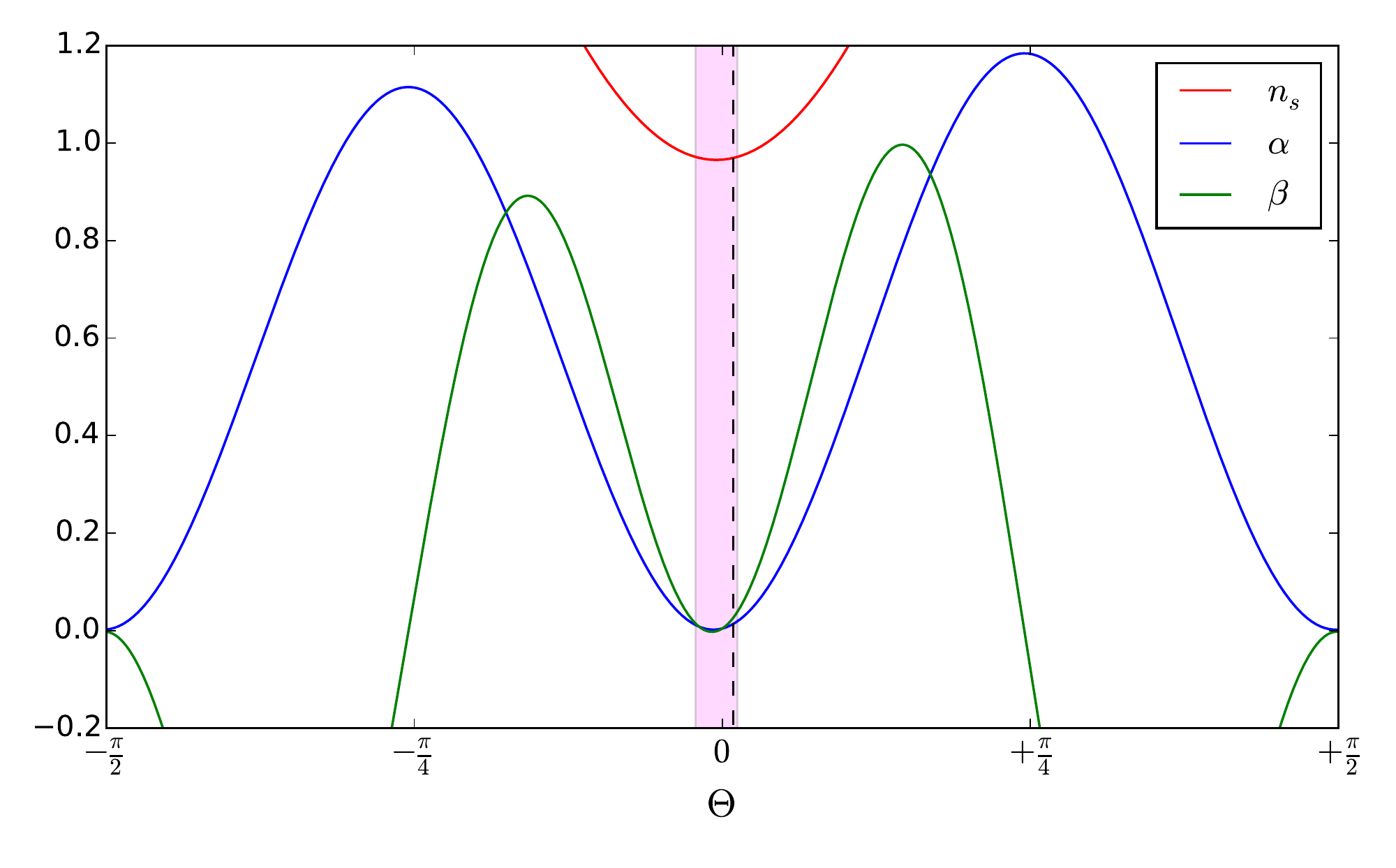}
    \caption{Predictions for the case (\ref{casetwo}), with $m/M_{\rm Pl} = 5M/M_{\rm Pl} = g/5 = 4.8 \times 10^{-4}$ and $\xi = -0.125/M_{\rm Pl}$. $\Theta$ is calculated to be $2.8 \times 10^{-2}$, which is represented by the black dashed line and, as shown, this falls within the shaded region of favoured values for $n_s$. As in this model, $A^*$ and $B^*$ are larger than in the uncoupled case, the oscillations in $\alpha_s$ and $\beta_s$ as a function of $\Theta$ are amplified so a much wider range of phenomenology is clearly possible. For the transfer angle calculated, we obtain $n_s = 0.970$, $\alpha_s = 0.014$, and $\beta_s = 0.027$.}
    \label{fig:two}
\end{figure}

Let us now consider a more general choice with 
\beq \label{casetwo}
V =  \frac{1}{2} m^2 \phi^2 + \frac{1}{2} M^2 \chi^2 + \frac{1}{2} g^2 \phi^2 \chi^2 \, , \quad b = - \xi \phi \, ,
\eeq
which results in 
\begin{eqnarray}
A &\simeq&\frac{4}{F^2 (\phi^2+e^{-2\xi \phi}\chi^2)} \left[ (\xi(\lambda^2 + \phi^2)+2 e^{\xi \phi}\phi)\chi e^{-2\xi \phi} \dot{\chi}^2  \right. \nonumber\\
 &-& \left. (e^{2 \xi \phi}\mu^2 - \lambda^2 + e^{2\xi \phi}\chi^2 - \phi^2) e^{-\xi\phi} \dot{\phi}\dot{\chi} - 2 \phi \chi e^{\xi \phi} \dot{\phi}^2 \right] \, ,  
\end{eqnarray}
and
\begin{eqnarray} 
B &\simeq& - 2 \epsilon_0 + \frac{2 \xi (\lambda^2 + \phi^2) \chi}{F^2}  \nonumber \\
&+& \frac{2}{F^2 (\phi^2+e^{-2\xi \phi}\chi^2)}  \left[ ((1+\xi \chi)(\lambda^2+\phi^2)e^{2\xi \phi}) \right. \nonumber \\ 
&-& \left. (\mu^2+\chi^2) e^{-2 \xi \phi} \dot{\chi}^2  - (\xi (\lambda^2+\phi^2) - 8 e^{\xi \phi} \phi) \chi e^{- \xi\phi} \dot{\phi}\dot{\chi} \right. \nonumber \\
&+& \left. (\mu^2 - e^{2 \xi \phi} \lambda^2 + \chi^2 - e^{2 \xi \phi} \phi^2) \dot{\phi}^2 \right] \, ,
\end{eqnarray}
where $F^2 = (\mu^2 \phi^2 + (\lambda^2 + \phi^2) \chi^2)$, $\mu = m/g$ and $\lambda = M/g$. There are now two more parameters in this theory, $g$ and $\xi$, which allow us to have larger values for $A$ and $B$ at horizon crossing than in the previous example. In Fig \ref{fig:two} we present a choice of parameters which demonstrates explicitly that this model is able to predict large enough values for $A$ and $B$ at horizon crossing such that we can achieve a much more general range of hierarchies of runnings in this model. Furthermore, we compute a first approximation for $\Theta$ by numerically integrating eq. (\ref{eq:TRSintegral}), assuming (\ref{eq:A}) and (\ref{eq:B}), and use this in eqs. (\ref{eq:ns})--(\ref{eq:beta}) to make predictions for the spectral properties for this case. The computed value of $\Theta$ is shown in Fig \ref{fig:two} with a dashed black line, at which we obtain $\mathcal{P}_\mathcal{R} = 2.21 \times 10^{-9}$, $n_s = 0.970$, $\alpha_s = 0.014$, and $\beta_s = 0.027$.  We note that in this example, $\beta_s^* = -1 \times 10^{-3}$, confirming our expectations from the preceding discussion that the single-field-like spectrum at horizon crossing has small and negative $\beta_s$, but the superhorizon amplification by isocurvature effects is able produce results more consistent with \cite{Escudero:2015wba,Cabass:2016ldu}. At present, we give only this example to explicitly show that our approach is feasible, leaving comprehensive analyses of the running of the running in this model and other interesting models for future work. 

Note in Fig \ref{fig:two} that for $|\Theta| \approx \pi/2$ the runnings both approach $0$ due to the factor of $\cos \Theta$ in each of their expressions, but in this regime the spectrum is heavily blue-tilted as no such suppression occurs in $n_s$, which is proportional instead to $\sin \Theta$. For intermediate values of $\Theta$, the runnings oscillate and are generally not going to be of the right order simultaneously. One may somewhat generally expect (though perhaps not entirely excluding other possibilities) then, that if $B_*$ and $A_*$ are sufficiently large to make $\beta_s > \alpha_s$, one would need a model which predicts a small transfer angle as it is in this regime that the spectral index and the runnings can all simultaneously be of the right magnitude. This is consistent with the example we gave in Fig \ref{fig:two}, where $\Theta \approx 2.8 \times 10^{-2}$. Note that a small transfer angle and somewhat large $A_*$ and $B_*$ values are not necessarily contradictory, as while $A_*$ and $B_*$ only contain information from the moment of horizon crossing, $\Theta$ encodes the entire evolution of the perturbations from this point until the end of inflation.

In the regime where $\Theta$ is small, the leading order behaviour ($O(\Theta^0)$) of the spectral index and runnings is,
\begin{eqnarray} 
&n_s & \simeq n_s^* \,  \\
&\alpha_s& \simeq \alpha_s^* + 2 A_*^2 \, \\
&\beta_s& \simeq \beta_s^*  - 6 A_*^2 B_*  \, . 
\end{eqnarray}
From this we can infer some generally desirable properties for $A_*$ and $B_*$ in this limit. To have a positive amplification of $\beta_s$, $B_*$ should be negative and large. The sign of $A_*$ doesn't matter and the required magnitude is determined solely by the value of $\alpha_s$. Furthermore, from the consistency relation (\ref{eq:twofieldcr}), we can see that a small transfer angle would imply approximately that $r \simeq - 8 n_T$. Two-field models with small transfer angles hence, somewhat tantalizingly, predict a consistency relation almost indistinguishable from that of single-field inflation.

To conclude, in this paper we applied the transfer function formalism to derive the expressions for the running of the spectral index $\alpha_s$ and its running $\beta_s$ in general two--field inflationary scenarios. We find that entropy perturbations significantly affect not only the value of $n_s$, but also $\alpha_s$ and $\beta_s$, and this may be useful in explaining the recent hints of a large $\beta_s$ that have appeared in the literature. Should this observation be confirmed to a higher statistical significance in future CMB experiments, or even if, more pessimistically, it is later found that $\beta_s$ appears to be of a similar magnitude to $\alpha_s$, this may serve as a powerful discriminator between models of cosmology. In particular, we have argued here that single field models and non-interacting two field models are naturally not capable of explaining positive runnings with a hierarchy such as $\beta_s > \alpha_s$. Slow-roll violating models and those with non-trivial evolutions of sound speeds, such as those with spectral features, could be able to provide exceptions to these arguments, but may require considerable fine-tuning to generate precisely this kind of hierarchy. 

By introducing kinetic interactions between two scalar fields, we were able to show that is it feasible that such models can produce a large running of the running and gave a specific example in which this is realised in a way which is largely consistent with the analyses in \cite{Escudero:2015wba,Cabass:2016ldu}. While it is left to future work to comprehensively study individual models of inflation and categorise their predictions of the running and its running, in this letter we have made initial exploratory steps towards this goal, and shown that some of the simplest models such as single-field and non-interacting two-field models would be difficult to reconcile with evidence of a large positive $\beta_s$, but a smaller or similar magnitude $\alpha_s$.

We hence argue that the confirmation of this hint of interesting runnings would provide a strong motivation for the study of extended models of inflation, which could in turn tell us a lot about the physics of the early universe. Future CMB experiments would be of great value in facilitating this approach, particularly those such as PIXIE, whose data on the small scales probed by measurements of spectral distortions should help strengthen our constraints on thus-far weakly probed parameters like $\alpha_s$ and $\beta_s$ \cite{Cabass:2016ldu,Chluba:2015bqa,Chluba:2016bvg,Cabass:2016giw}. Use of the consistency relation for two-field inflation along with observations of the primordial tensor power spectrum will also be a valuable tool in testing these kinds of models, and hence proposed missions such as PRISM \cite{Andre:2013nfa,Andre:2013afa} would also directly benefit work in this direction.

\begin{acknowledgments}
The work of CvdB is supported by the Lancaster- Manchester-Sheffield Consortium for Fundamental Physics under STFC Grant No. ST/L000520/1. CL is supported by a STFC studentship. 
\end{acknowledgments}

\bibliography{Running_refs}

\end{document}